\documentclass[12pt]{article}

\begin{document}

\begin{titlepage}

\begin{flushright}
IUHET-503\\
\end{flushright}
\vskip 2.5cm

\begin{center}
{\Large \bf Bounds on Spin-Dependent Lorentz Violation\\
From Inverse Compton Observations}
\end{center}

\vspace{1ex}

\begin{center}
{\large Brett Altschul\footnote{{\tt baltschu@indiana.edu}}}

\vspace{5mm}
{\sl Department of Physics} \\
{\sl Indiana University} \\
{\sl Bloomington, IN 47405 USA} \\

\end{center}

\vspace{2.5ex}

\medskip

\centerline {\bf Abstract}

\bigskip

Some of the best bounds on possible Lorentz violation in the electron sector come
from observations of high-energy astrophysical phenomena. Using measurements of
TeV inverse Compton radiation from a number of sources, we place the first
bounds---at the $10^{-15}$ level---on seven of the electron $d$ coefficients.

\bigskip

\end{titlepage}

\newpage

In the last ten years, there has been growing interest in the possibility that
special relativity may not be exactly correct. Lorentz invariance might be only
approximately valid, with there existing small deviations from rotation and boost
invariance. A variety of experimental tests have been performed, constraining
many types of Lorentz violations to be small. However, many other quite reasonable
forms of Lorentz violation are poorly, if at all, constrained.

In quantum field theory language, Lorentz violation is described by the standard
model extension (SME)~\cite{ref-kost1,ref-kost2}. Lorentz-violating corrections
to gravity can also be incorporated~\cite{ref-kost12}. The minimal SME, which
contains gauge invariant, renormalizable parameters, provides a convenient
parameterization of possible Lorentz violations, in terms of various background
tensor coefficients. There are separate sets of coefficients for each species of
particle in the theory.

In this paper, we are specifically interested in electron Lorentz violation.
Bounds on the Lorentz-violating coefficients for electrons come from clock
comparison experiments~\cite{ref-berglund,ref-kost6}, experiments with spin-polarized
matter~\cite{ref-hou,ref-heckel2}, Michelson-Morley experiments with cryogenic
resonators~\cite{ref-antonini,ref-herrmann,ref-stanwix,ref-muller2}, and Doppler
effect measurements~\cite{ref-saathoff,ref-lane1}.
Finally, some of the best bounds come from high-energy
astrophysics~\cite{ref-stecker,ref-jacobson1,ref-jacobson2,ref-altschul6,
ref-altschul7}. A number of astrophysical processes have been used to set these
bounds. We shall show that observations of one of these processes, inverse Compton
(IC) scattering, can be used to place tight bounds on seven SME parameters that have
not been bounded before.

The minimal SME Lagrange density for the electron sector is
\begin{equation}
\label{eq-L}
{\cal L}=\bar{\psi}(i\Gamma^{\mu}\partial_{\mu}-M)\psi,
\end{equation}
where
\begin{equation}
M=m+\!\not\!a-\!\not\!b\gamma_{5}+\frac{1}{2}H^{\mu\nu}\sigma_{\mu\nu}+im_{5}
\gamma_{5},
\end{equation}
and
\begin{equation}
\Gamma^{\mu}=\gamma^{\mu}+c^{\nu\mu}\gamma_{\nu}-d^{\nu\mu}\gamma_{\nu}
\gamma_{5}+e^{\mu}+if^{\mu}\gamma_{5}+\frac{1}{2}g^{\lambda\nu\mu}
\sigma_{\lambda\nu}.
\end{equation}
However, some of the coefficients, such as $f$ and $a$, are
unphysical~\cite{ref-altschul8}.
At high energies, the Lorentz-violating effects coming from $\Gamma^{\mu}$ are
more important than those coming from $M$, simply because the $\Gamma^{\mu}$ is
multiplied by a factor of the momentum. The coefficients contained in $M$ do not
affect the maximum achievable velocity (MAV) of the particles, but the ones included
in $\Gamma^{\mu}$ do.

In order to canonically quantize the electron field, we must have $\Gamma^{0}=
\gamma^{0}$ in the frame where the quantization is to be performed. Otherwise,
there are nonstandard time derivatives that make defining a Hamiltonian
impossible. If all Lorentz violations are small, it is possible to perform a
rotation in spinor space so that in the new basis, $\Gamma^{0}$ does not contain
any unconventional terms. Then, in the frame in which the theory is quantized,
we must have $c^{\nu0}=d^{\nu0}=e^{0}=g^{\lambda\nu0}=0$. We shall adopt this
convention here, choosing as the quantization frame the rest frame of the sun
and using sun-centered celestial equatorial coordinates
$(X,Y,Z,T)$~\cite{ref-bluhm4}.

In the full SME, the $e$ and $g$ terms are actually forbidden if we demand
renormalizability. These terms mix left- and right-handed leptons and hence break
$SU(2)_{L}$ gauge invariance. 
This gauge invariance must be present prior to
spontaneous symmetry breaking if the theory is to be renormalizable; so $e$ and $g$
could only arise as part of the spontaneous breaking.
However,
the operators parameterized by $e$ and $g$ are of dimension four, so they
could only appear as vacuum expectation values of higher-dimension, nonrenormalizable
operators. Even if we do
not insist on renormalizability, we would expect nonrenormalizable terms like $e$
and $g$ to be suppressed relative to the renormalizable ones.

The renormalizable operators contained in $\Gamma^{\mu}$ are described by $c$ and
$d$. The $c$ terms and $d$ terms are similar in form, although the former are
obviously simpler. There are already some fairly good bounds on the $c$ terms in
the electron sector, most of them coming from astrophysical data; however, the $d$
terms have been less explored. At nonrelativistic energies, the effects of $d$
must enter in specific combinations of $d$ and $H$, and all existing bounds on $d$
are actually bounds on these particular combinations. However, using relativistic
tests, bounds on $d$ may be obtained separately from $H$.

Our focus in this paper
will be on the $d$ terms and how they affect the MAV of electrons. 
The effects of
$c$ have previously been considered~\cite{ref-altschul6,ref-altschul7}; with a $c$
term only, the maximum electron velocity in a direction $\hat{e}$ is
\begin{equation}
\left(v_{j}\hat{e}_{j}\right)_{\max}=1-c_{jk}\hat{e}_{j}\hat{e}_{k}-c_{0j}
\hat{e}_{j}.
\end{equation}
If this is less than one in a given direction, there will be a maximum Lorentz
factor $\gamma=\left(1-\vec{v}\,^{2}\right)^{-1/2}$ for particles moving in that
direction, which would have observable effects on the synchrotron spectra of
high-energy sources. On the other hand, if 
$\left(v_{j}\hat{e}_{j}\right)_{\max}$ were greater than one, there would
exist a finite maximum energy for particles with speeds less than one. More
energetic particles, moving faster than the speed of light, would lose energy quickly
through vacuum Cerenkov radiation. Lower limits on electrons' maximum subluminal
energies can be inferred from sources' IC spectra. Combining
the synchrotron and IC information for a single source, we can obtain a two-sided
bound on a particular linear combination of the $c^{\nu\mu}$ coefficients.

By observing the radiation from electrons of energy $E$, we could place bounds on
$c$ that are ${\cal O}(m^{2}/E^{2})$. This energy scale at which phenomena such
as vacuum Cerenkov radiation and single-photon pair creation would appear is
$m/\sqrt{|c|}$~\cite{ref-kost3}, and so the absence of these effects
up to a scale $E$ indicates $|c|$ must be smaller than the indicated
${\cal O}(m^{2}/E^{2})$. The forms of Lorentz violation parameterized by $c$ are
especially simple; the effects of $c$ are independent of particle spin, and
the $c$ term in the Lagrangian is even under both $C$ and $PT$.

The situation with a $d$ instead of a $c$ is trickier. The $d$ term in
$\Gamma^{\mu}$ differs only from the $c$ term by the presence of a $\gamma_{5}$, but
it is odd under $C$ and $PT$. At
ultrarelativistic energies where the mass can be neglected, $\gamma_{5}$ is the
helicity operator and can be diagonalized simultaneously with the Hamiltonian.
So at these high energies, $d$ resembles a $c$ whose sign depends on the helicity.
Therefore, the sign of contribution of $d$ to the MAV is helicity dependent, and
hence there can never be a MAV less than one. (One helicity will have a MAV less than
one, but the other will not. In real sources, the spins are not polarized, and so the
physically relevant MAV must be a maximum over all spin states.)
However, it turns out that IC data can
actually give two-sided bounds on $d$ by constraining the possibility that the MAV
may be greater than one, even though the same kinds of data only give one-sided
bounds on $c$.

With a $d$ term only, we would expect that the MAV in a given direction would
become
\begin{equation}
\left(v_{j}\hat{e}_{j}\right)_{\max}=1+sd_{jk}\hat{e}_{j}\hat{e}_{k}+sd_{0j}
\hat{e}_{j},
\end{equation}
where $s$ is now the helicity, just based on the similar way that $c$ and $d$
enter into ${\cal L}$.
However, $\gamma_{5}$ is strictly only the helicity operator when the electron mass
vanishes. The mass
will enter unavoidably into our bounds, and this raises the question
of whether simply making the replacement $\gamma_{5}\rightarrow s$ is really valid.
This is most easily answered by looking at the explicit energy-momentum relation
and group velocity in the presence of both $d$ and $m$; this has been done
in~\cite{ref-altschul4}. The dispersion relation is simplest if only
$d_{0j}$ is nonzero, when we have
\begin{equation}
E^{2}=m^{2}+\left(\left|\vec{\pi}\,\right|+sd_{0j}\pi_{j}\right)^{2}
\end{equation}
and
\begin{equation}
\left(v_{g}\right)_{k}=\frac{\left|\vec{\pi}\,\right|+sd_{0j}\pi_{j}}{E}\left(\hat
{\pi}_{k}+sd_{0k}\right).
\end{equation}
The maximum speed for a given helicity $s$ is, to leading order in $d$, $1+sd_{0j}
\hat{\pi}_{j}$, and the velocity is superluminal if
\begin{equation}
\label{eq-d0limit}
sd_{0j}\hat{\pi}_{j}<-\frac{1}{2(E/m)^{2}}.
\end{equation}
The energy-momentum relation with $d_{jk}$ is substantially more cumbersome,
but the pattern is the same. The net result is that $v>1$ whenever
\begin{equation}
\label{eq-dlimit}
sd_{0j}\hat{\pi}_{j}+sd_{jk}\hat{\pi}_{j}\hat{\pi}_{k}<-\frac{1}{2(E/m)^{2}}.
\end{equation}

It is obvious that by choosing one or the other sign for $s$, we can ensure that
the inequality (\ref{eq-dlimit}) does not hold and so the electron is moving more
slowly than light. Thus there is no true maximum subluminal energy if we consider
both spin states.  Nonetheless, because the helicity is not a constant of
motion, we can extract bounds from the IC data.

It is natural to expect that in an astrophysical source, energy will be roughly
evenly distributed between left- and right-handed electrons. Yet we may worry
that because of the differing kinematics of the two helicities, the energy may not be
so evenly distributed. However, even if the initial distribution of electrons were
completely polarized at the highest energies (with only one helicity state
occupied), the presence of a magnetic field in a source will destroy this
polarization. Because an electron has an anomalous magnetic moment, the spin
will precess during cyclotron motion, changing by
${\cal O}(e^{2}\gamma)$ with each
revolution. Moreover, since the MAV is direction- as well as spin-dependent, a
particle with fixed helicity and energy may be subluminal when moving in certain
directions along its trajectory and superluminal when moving in other directions.
So if we can infer the presence of electrons of a given energy in a source, there
must be electrons of that energy with both helicities. Further, if there is no
evidence of vacuum Cerenkov radiation, then both helicities at that energy must
be moving more slowly than light.

If we observe IC photons with energies up to $E_{\max}$ coming at us from a source
along the direction $\hat{e}$, there must be electrons equally energetic in the
source, also moving in the $\hat{e}$ direction. If the radiation from the source is
well understood, and vacuum Cerenkov radiation is not present, then it follows that
\begin{equation}
\label{eq-2sidedlimit}
\left|d_{0j}\hat{e}_{j}+d_{jk}\hat{e}_{j}\hat{e}_{k}\right|<\frac{1}{2(E_{\max}/m)
^{2}}.
\end{equation}

If we were to include the effects of $c$, we would find
\begin{equation}
\label{eq-boundwithc}
-c_{0j}\hat{e}_{j}-c_{jk}\hat{e}_{j}\hat{e}_{k}+
\left|d_{0j}\hat{e}_{j}+d_{jk}\hat{e}_{j}\hat{e}_{k}\right|<\frac{1}{2(E_{\max}/m)
^{2}}.
\end{equation}
(It immediately follows that including the effects of $d$ would only strengthen
the previously derived IC bounds on $c$.)
However, there are good reasons to believe that the magnitude of $c$ is at least
as small as the bounds we shall be placing on $d$ here. The synchrotron spectrum
is less sensitive to $d$ than to $c$, and although the synchrotron data only give
one-sided bounds on $c$, they are suggestive the overall order of magnitude that $c$
may be. Even stronger restrictions on $c$ come from naturalness conditions, since
there are much stronger bounds on $c$---at the $10^{-25}$ level---for the
proton~\cite{ref-wolf}.
Radiative corrections mix the $c$ coefficients for different charged particles, so
it would be highly unnatural for the electron $c$ to be many orders of magnitude
larger than the proton $c$. We shall therefore assume than $c$ can be neglected
compared to $d$ in what follows.

Direct experimental bounds on the $d$ coefficients in the electron and proton
sectors are not nearly as good. In our formulation with $d_{\nu T}=0$,
there are bounds only on $d_{TX}$ and $d_{TY}$, and these are mixed with
bounds on $H$. However, the bounds in the electron sector are at the $10^{-19}$
level~\cite{ref-kost6},
which is substantially better than we can achieve with the IC data. Assuming that
there is no special cancellation between $d$ and $H$, a reasonably conservative
interpretation of these bounds is that $|d_{TX}|$ and $|d_{TY}|$ must be less
than $10^{-18}$. (In earlier formulations where the fermions were not quantized in
the sun-centered frame, and in which $d_{\nu T}$ was therefore nonvanishing, bounds
on $d_{T\nu}$ became entangled with bounds on $d_{\nu T}$ and $b_{\nu}$. Here, we
have avoided this complication by choosing the formulation of the theory that has the
most concise expression when the effects of $d$ are predominant.)

Bounds on an electron MAV less than
one have also been derived from looking at the threshold for the pair creation
process $\gamma\rightarrow e^{+}+e^{-}$, which is forbidden if Lorentz symmetry is
exact. This can provide excellent bounds on a $c$ term, but not on a $d$. The bounds
on $c$ come from observations of TeV photons from sources such as the Crab nebula.
If the MAV for both electron spin states is less than one, then a very energetic
photon may decay into two electrons, with a decay rate of ${\cal O}(e^{2})$. If we
observe photons from distant sources up to a given energy, we know that this decay
is not possible up to at least the measured energy. However, the situation is not
the same with $d$, because of angular momentum complications. All the particles in
the process $\gamma\rightarrow e^{+}+e^{-}$ are nearly collinear, so if angular
momentum is conserved (which it nearly is), the helicities of all three particles
must be the same---a right-handed photon decays to two right-handed electrons, and
the same with left-handed particles. However, because the
electron MAV with $d$ depends on the helicity,
the process will be kinematically allowed only for one of the two helicity options
outlined above. If the process is allowed for right-handed quanta, it is not allowed
for left-handed ones, and so some half of the photons coming from a given source
will not decay into electron-positron pairs. (Although angular-momentum is not
exactly conserved in Lorentz-violating theories, it is still approximately
conserved. Decays that do not conserve angular momentum will be suppressed by two
factors of the Lorentz-violating coefficient $d$, since $d$ would need to appear
explicitly in the amplitude for the process. Because $d$ is expected to be miniscule,
such decays should be quite slow to happen.)

\begin{table}
\begin{center}
\begin{tabular}{|l|c|c|c|c|}
\hline
Emission source & $\hat{e}_{X}$ & $\hat{e}_{Y}$ & $\hat{e}_{Z}$ & $E_{\max}/m$ \\
\hline
Crab nebula & $-0.10$ & $-0.92$ & $-0.37$ &
$2\times 10^{8}$\cite{ref-tanimori} \\
G 0.9+0.1 & 0.05 & 0.88 & 0.47 &
$10^{7}$\cite{ref-aharonian9} \\
G 12.82-0.02 & $-0.06$ & 0.95 & 0.29 &
$5\times 10^{7}$\cite{ref-aharonian6} \\
G 18.0-0.7 & $-0.11$ & 0.97 & 0.24 &
$7\times 10^{7}$\cite{ref-aharonian8,ref-aharonian11} \\
G 347.3-0.5 & 0.16 & 0.75 & 0.64 &
$2\times 10^{7*}$\cite{ref-aharonian3} \\
MSH 15-52 & 0.34 & 0.38 & 0.86 &
$8\times 10^{7}$\cite{ref-aharonian4} \\
Mkn 421 & 0.76  & $-0.19$  & $-0.62$ &
$3\times 10^{7*}$\cite{ref-albert,ref-aharonian7} \\
Mkn 501 & 0.22 &0.74 & $-0.64$ &
$4\times 10^{7*}$\cite{ref-aharonian10} \\
PSR B1259-63 & 0.42 & 0.12 & 0.90 &
$6\times 10^{6*}$\cite{ref-aharonian5} \\
SNR 1006 AD & 0.52  & 0.53 & 0.67 &
$7\times 10^{6}$\cite{ref-allen} \\
Vela SNR & 0.44 & $-0.55$ & 0.71 &
$1.3\times 10^{8}$\cite{ref-aharonian2} \\
\hline
\end{tabular}
\caption{
\label{table-IC}
Parameters for the IC sources that we shall use to constrain $d$.
References are given for each value of $E_{\max}$.
The four $E_{\max}/m$ values marked with asterisks denote the four sources for which
the IC origin of the observed $\gamma$-rays is merely strongly favored, rather than
completely assured.}
\end{center}
\end{table}

So an analysis of IC data appears to offer the best prospects for bounding $d$. We
have previously collected a number of data points from the observational literature.
These are replicated in table~\ref{table-IC}. The data come from sources of TeV
IC photons. (In a few cases, it is not completely certain---although it does appear
strongly favored---that the high-energy spectrum
in entirely due to IC emission.) $E_{\max}$ is the highest observed photon energy
from each of these sources. These sources all have well-understood spectra, with no
indications of the vacuum Cerenkov radiation we would expect to see if there were
superluminal electrons. Therefore, each $E_{\max}$ is a lower bound on the maximum
subluminal energy for electrons moving in the source-to-Earth direction $\hat{e}$.
The components of each $\hat{e}$ are given in the sun-centered coordinate system.

\begin{table}
\begin{center}
\begin{tabular}{|c|c|c|}
\hline
$|d_{\nu\mu}|$ & Bound (no priors) & Bound (with priors) \\
\hline
$|d_{XX}|$ & $8\times 10^{-14}$ & $2\times 10^{-14}$ \\
$|d_{YY}|$ & $7\times 10^{-15}$ & $3\times 10^{-15}$ \\
$|d_{ZZ}|$ & $2\times 10^{-14}$ & $3\times 10^{-15}$ \\
$|d_{(XY)}|$ & $5\times 10^{-14}$ & $2\times 10^{-15}$ \\
$|d_{(YZ)}|$ & $7\times 10^{-14}$ & $2\times 10^{-14}$ \\
$|d_{(YZ)}|$ & $2\times 10^{-14}$ & $7\times 10^{-15}$\\
$|d_{TX}|$ & $5\times 10^{-14}$ & - \\
$|d_{TY}|$ & $5\times 10^{-15}$ & - \\
$|d_{TZ}|$ & $4\times 10^{-16}$ & $8\times 10^{-17}$\\
\hline
\end{tabular}
\caption{
\label{table-separate}
Independent bounds on the components of $d$, as determined both with and without the
inclusion of the earlier bounds on $|d_{TX}|$ and $|d_{TY}|$ coming from clock
comparison experiments~\cite{ref-kost6}.}
\end{center}
\end{table}

The data from table~\ref{table-IC}, when combined with
inequality~(\ref{eq-2sidedlimit}), give eleven two-sided bounds on various
combinations of the $d$ coefficients. This is enough to bound each of the six
$d_{(jk)}=d_{jk}+d_{kj}$ and the three $d_{Tj}$ coefficients above and below.
The bounds of the form (\ref{eq-2sidedlimit}) may be translated into bounds on
the separate coefficients by means of linear programming. The results are shown in
table~\ref{table-separate}. Unlike with previous results for $c$, where synchrotron
and IC data points each provided half of a two-sided inequality, both the
upper and lower limits on each combination of $d$ coefficients come from a single
data point. Because of this, the separate bounds on the various $d_{\nu\mu}$ are
all symmetric about zero.

The first column of numbers in table~\ref{table-separate} contains the bounds coming
solely from the IC data. These are the absolute maximum values that each
$|d_{\nu\mu}|$ may take when the components are subject to all eleven of the
inequalities (\ref{eq-2sidedlimit}). The second set of bounds were calculated using
the same method, but this time the linear program also incorporated priors from the
clock comparison data. The clock comparison bounds on $d_{TX}$ and $d_{TY}$ are
several orders of magnitude better than the astrophysical bounds, so the second
set of bounds are effectively just the limits on the various $|d_{\nu\mu}|$ when
$d_{TX}$ and $d_{TY}$ are set to zero. Including these priors improves the bounds
on the other coefficients by factors of two to twenty-five.
The bounds on $d$, with the clock comparison priors, are roughly comparable to the
best bounds on $c$ (which are mostly astrophysical, but also incorporate some
different laboratory results). This is not surprising, since the tightness of each
set of bounds is determined by the highest electron energies in the sources we can
observe. If it turns out that the magnitude of $c$ is actually right at the
$10^{-15}$ level implied by the best bounds, then including the effects of $c$
according to (\ref{eq-boundwithc}) would worsen the bounds on $d$ slightly, but by
no more than an ${\cal O}(1)$ factor.

The electron $c$ coefficients have been bounded
using analyses of synchrotron spectra, analyses of IC spectra, and the observed
absence of the process $\gamma\rightarrow e^{+}+e^{-}$.
The scales of these bounds are
all roughly similar, as they are all ultimately determined by the same basic
quantity, which is the maximum energy attained by individual particles in energetic
sources.
However, the $d$ terms
have been harder to constrain. Of the astrophysical data mentioned above, only the
IC data really provide good bounds on $d$. Our new IC bounds on $d$ include bounds on
seven Lorentz-violating coefficients that have not previously been bounded, at a
$10^{-15}$ level comparable to the best astrophysical bounds on $c$.

\section*{Acknowledgments}
The author is grateful to V. A. Kosteleck\'{y} for helpful discussions.
This work is supported in part by funds provided by the U. S.
Department of Energy (D.O.E.) under cooperative research agreement
DE-FG02-91ER40661.


\begin{thebibliography}{99}

\bibitem{ref-kost1}D. Colladay, V. A. Kosteleck\'{y}, Phys. Rev. D {\bf 55},
6760 (1997).
\bibitem{ref-kost2}D. Colladay, V. A. Kosteleck\'{y}, Phys. Rev. D {\bf 58},
116002 (1998).
\bibitem{ref-kost12}V. A. Kosteleck\'{y}, Phys. Rev. D {\bf 69}, 105009 (2004).

\bibitem{ref-berglund}C. J. Berglund, L. R. Hunter, D. Krause, Jr., E. O.
Prigge, M. S. Ronfeldt, S. K. Lamoreaux, Phys. Rev. Lett. {\bf 75}, 1879 (1995).
\bibitem{ref-kost6}V. A. Kosteleck\'{y}, C. D. Lane, Phys. Rev. D {\bf 60},
116010 (1999).
\bibitem{ref-hou}L.-S. Hou, W.-T. Ni, Y.-C. M. Li, Phys. Rev. Lett. {\bf 90}, 201101
(2003).
\bibitem{ref-heckel2}B. R. Heckel, C. E. Cramer, T. S. Cook, S. Schlamminger,
E. G. Adelberger, U. Schmidt, Phys. Rev. Lett. {\bf 97}, 021603 (2006).
\bibitem{ref-antonini}P. Antonini, M. Okhapkin, E. G\"{o}kl\"{u}, S. Schiller, Phys.
Rev. A {\bf 71}, 050101 (2005).
\bibitem{ref-stanwix}P. L. Stanwix, M. E. Tobar, P. Wolf, M. Susli, C. R. Locke,
E. N. Ivanov, J. Winterflood, F. van Kann, Phys. Rev. Lett. {\bf 95}, 040404
(2005).
\bibitem{ref-herrmann}S. Herrmann, A. Senger, E. Kovalchuk, H. M\"{u}ller, A. Peters,
Phys. Rev. Lett. {\bf 95}, 150401 (2005).
\bibitem{ref-muller2}H. M\"{u}ller, Phys. Rev. D {\bf 71}, 045004 (2005).
\bibitem{ref-saathoff}G. Saathoff, S. Karpuk, U. Eisenbarth, G. Huber, S. Krohn, R.
Mu\~{n}oz Horta, S. Reinhardt, D. Schwalm, A. Wolf, G. Gwinner, Phys. Rev. Lett.
{\bf 91}, 190403 (2003).
\bibitem{ref-lane1}C. D. Lane, Phys. Rev. D {\bf 72}, 016005 (2005).

\bibitem{ref-stecker}F. W.  Stecker, S. L.  Glashow, Astropart. Phys. {\bf 16},  97
(2001).
\bibitem{ref-jacobson1}T. Jacobson, S. Liberati, D. Mattingly, Nature {\bf 424},
1019 (2003).
\bibitem{ref-jacobson2}T. Jacobson, S. Liberati, D. Mattingly, F. W. Stecker,
Phys. Rev. Lett. {\bf 93}, 021101 (2004).
\bibitem{ref-altschul6}B. Altschul, Phys. Rev. Lett. {\bf 96}, 201101 (2006).
\bibitem{ref-altschul7}B. Altschul, Phys. Rev. D {\bf 74}, 083003 (2006).
\bibitem{ref-altschul8}B. Altschul, J. Phys. A {\bf 39}, 13757 (2006).
\bibitem{ref-bluhm4}R. Bluhm, V. A. Kosteleck\'{y}, C. D. Lane, N. Russell, Phys.
Rev. D {\bf 68}, 125008 (2003).
\bibitem{ref-kost3}V. A. Kosteleck\'{y}, R. Lehnert, Phys. Rev. D {\bf 63},
065008 (2001).
\bibitem{ref-altschul4}B. Altschul, D. Colladay, Phys. Rev. D {\bf 71}, 125015
(2005).
\bibitem{ref-wolf}P. Wolf, F. Chapelet, S. Bize, A. Clairon,  Phys. Rev. Lett.
{\bf 96}, 060801 (2006).
\bibitem{ref-tanimori}T. Tanimori, {\em et al.}, 1998, Astrophys. J. {\bf 492}, L33
(1998).
\bibitem{ref-aharonian9}F. A. Aharonian, {\em et al.}, Astron. Astrophys. {\bf 432},
L25 (2005).
\bibitem{ref-aharonian6}F. A. Aharonian, {\em et al.}, Astrophys. J. {\bf 636}, 777
(2006).
\bibitem{ref-aharonian8}F. A. Aharonian, {\em et al.}, Astron. Astrophys. {\bf 442},
L25 (2005).
\bibitem{ref-aharonian11}F. A. Aharonian, {\em et al.}, astro-ph/0607548.
\bibitem{ref-aharonian3}F. A. Aharonian, {\em et al.}, Nature {\bf 432}, 75 (2004).
\bibitem{ref-aharonian4}F. A. Aharonian, {\em et al.}, Astron. Astrophys. {\bf 435},
L17 (2005).
\bibitem{ref-albert}J. Albert, {\em et al.}, astro-ph/0603478.
\bibitem{ref-aharonian7}F. A. Aharonian, {\em et al.}, Astron. Astrophys. {\bf 437},
95 (2005).
\bibitem{ref-aharonian10}F. A. Aharonian, {\em et al.}, Astron. Astrophys. {\bf 349},
11 (1999).
\bibitem{ref-aharonian5}F. A. Aharonian, {\em et al.}, Astron. Astrophys. {\bf 442},
1 (2005).
\bibitem{ref-allen}G. E. Allen, R. Petre, E. V. Gotthelf, Astrophys. J. {\bf 558},
739 (2001).
\bibitem{ref-aharonian2}F. A. Aharonian, {\em et al.}, Astron. Astrophys. {\bf 448},
L43 (2006).

\end{thebibliography}
\end{document}